\documentclass[a4paper,12pt]{article} 

\usepackage{ulem}
\usepackage{geometry,epsfig,subfigure}
\usepackage{lineno}

\usepackage{amsmath}
\usepackage{amssymb}
\usepackage{graphics,epsfig}

\usepackage[square,sort&compress,comma,numbers]{natbib}

\usepackage[usenames]{color}
\usepackage{fancyhdr,url}   
\usepackage{setspace}  
\usepackage{multirow}   
\usepackage{rotating}    
\usepackage{colortbl}     
\usepackage{relsize} 
\usepackage{booktabs}
\usepackage{cancel}
\usepackage{enumerate}
\numberwithin{equation}{section}
\usepackage{soul} 
\usepackage{xcolor} 
\usepackage{pdflscape}
\usepackage{hyperref}
\usepackage[capitalise]{cleveref}
\usepackage{lineno}
\usepackage{graphicx}
\usepackage{etoolbox}
\usepackage{epstopdf}

\definecolor{darkblue}{rgb}{0,0,0.8}
\definecolor{darkgreen}{rgb}{0,0.5,0}

\long\def\symbolfootnote[#1]#2{\begingroup \def\thefootnote{\fnsymbol{footnote}}\footnote[#1]{#2} \endgroup} 

\newcommand{\del}{\ensuremath{\partial}}

\renewcommand{\cos}[1]{ \text{cos}\hspace{0.0cm}\left( {#1} \right) }

\newcommand{\degree}{\ensuremath{ ^{\circ}  }}

\usepackage{amsthm}

\newcommand{\HRule}{\rule{0.9\linewidth}{0.2mm}}

\begin{document}
\renewcommand*{\thepage}{\arabic{page}}

\setstretch{1.3}

\begin{center}
\large

\textbf{Interplay of capillary and Marangoni flows\\in micropillar evaporation}

\normalsize
\vspace{0.2cm}
Goksel Yuncu$^{a,b}$, Yigit Akkus$^{a}$, Zafer Dursunkaya$^{b}$\\
\smaller
\vspace{0.2cm}
$^a$ASELSAN Inc., 06200 Yenimahalle, Ankara, Turkey\\
$^b$Department of Mechanical Engineering, Middle East Technical University, 06800 \c Cankaya, Ankara, Turkey\
\vspace{0.2cm}
\end{center}

\begin{center} \noindent \HRule \\ \end{center}
\vspace{-0.6cm}
\begin{abstract}

\noindent  The evaporation from a micropillar evaporator is a problem governed by various interfacial phenomena such as the capillarity-induced liquid flow, thin-film evaporation intensifying near the contact lines, and thermocapillarity-induced Marangoni flow. However, past research has not been able to assess the effect of Marangoni flow due to the missing coupling between cell- and device-level modeling. In this work, we develop a comprehensive model for the evaporation from a micropillar evaporator by coupling the liquid flow with the energy transfer in both liquid and solid domains at both cell- and device-levels. The model is successfully validated with previous experiments. When the Marangoni number is sufficiently high, the model identifies a sharp reduction in the evaporator temperature attributed to the thermocapillary convection creating circulations beneath the liquid-vapor interface, which results in the formation of periodic reverse surface flows on the interface. This temperature drop cannot be identified when thermocapillarity is switched off in the model and the model's prediction substantially deviates from experimental measurement. Therefore, the current study reveals a hitherto unexplored role of Marangoni flow in the evaporation of water from micropillar evaporators. 

\vspace{0.2cm}

\end{abstract}
\vspace{-0.6cm}
\begin{center} \noindent \HRule \\ \end{center}

\pagebreak

\section{Introduction}
\label{sec:intro}

Continuous improvements in semi-conductor fabrication techniques have opened up an avenue for producing integrated circuits with increasing computational power in the past decades. While the decrease of feature size on a 2-D chip had determined the performance in conjunction with Moore's law \cite{moore1965}; currently, in the post-Moore era, performance is still enhanced by stacking 3-D chips vertically. Nevertheless, all this progress comes with an inevitable cost: excessive heat fluxes. Specifically, excessive local heat fluxes (\textit{e.g.} power amplifier hot spots exceeding \mbox{10$\, \rm kW \, cm^{-2}$} heat flux \cite{barcohen2021}) create severe thermal challenges, which are impossible to be handled by traditional approaches. These hot spots may be actively cooled down \textit{in situ} by intra-chip liquid circulation by establishing proper microfluidics-based solutions \cite{barcohen2021}. When intra-chip active liquid cooling could not be established, the heat from the hot spots should be transported to a wider heat removal area with a sufficiently small thermal resistance to eliminate the excessive temperature rise. 

When conduction is the sole energy transport mechanism, thermal resistance is determined by the thermal conductivity of the solid, which is inadequate to materialize the desired resistance values. This issue can be overcome by two-phase heat spreaders using the liquid-vapor phase change as the energy transport mechanism. The bottleneck in the design of two-phase heat spreaders is the evaporator due to the need for removing localized heat inputs of excessively large magnitudes without any dryout. The design success is determined based on optimizing the wick structure, which should provide sufficient liquid pumping together with minimum film resistance. A thorough optimization demands control over both the design parameters and fabrication. Owing to the advances in micro and nanofabrication techniques, the high-resolution control of surface topography over length scales ranging from molecular-level to macro-level becomes possible, which allows numerous opportunities for the heat transfer enhancement \textit{via} varying scale surface structures \cite{li2012enhancing,attinger2014,akkus2019first}. As opposed to the coating-based methods (\textit{e.g.} nanowire coating, CNT coating, etc.), which, in general, generate a random distribution of surface features, lithography-based methods are able to create features (\textit{e.g.} micro-scale and nano-scale posts, etc.) with exact shapes and sizes. Accordingly, thin-film evaporation from these engineered surfaces has gained substantial attention in recent decades.

Evaporators with micro-scale posts (commonly referred to as micropillars) have been extensively studied both experimentally \cite{coso2012enhanced,nam_2010, xiao_2010} and numerically \cite{ranjan_2012wicking,ravi_2014,zhu_2016}. Cylindrical micropillars attracted substantial attention among the different shaped micro-post evaporators due to their superior capillary pressure and permeability performance compared to their polygon-shaped counterparts with sharp corners \cite{ranjan_2012wicking}. The maximum dryout heat flux and the effective thermal resistance of the wick structure assess the performance of an evaporator  \cite{vaartstra_2019}. 
While the maximum dryout heat flux is proportional to the capillary pressure and permeability, thermal resistance scales down with the interfacial area and up with the film thickness. Therefore, accurate prediction of permeability, capillary pressure, and thermal resistance is essential for exploiting the full potential of an evaporator.

Many prior studies focused on the modeling of capillary liquid flow through the micropillar arrays without thermal considerations.  A common simplification was to assume a 2-D liquid flow through square and hexagonal packed arrays of infinitely long micropillars \cite{drummond_1984,gebart_1992,sobera_2006,tamayol_2011,yazdchi_2011}. The effect of interfacial geometry (\textit{i.e.} 3-D meniscus shape) was also negated in previous capillary flow models \cite{horner_2014,hale2014capillary,hale_2014_opt}.  These models primarily characterized the flow in terms of effective porosity and permeability that were obtained \textit{via} Darcy's law \cite{sobera_2006,tamayol_2011,horner_2014,yazdchi_2011} or Brinkman equation \cite{hale2014capillary,hale_2014_opt}. 

Because of its 3-D nature, the meniscus-shaped liquid-vapor interface established between the pillars cannot be straightforwardly determined. Surface Evolver \cite{SE_ref}, on the other hand, enables obtaining 3-D interfacial geometries based on the surface energy minimization principle. Many studies \cite{nam_2010,xiao_2010,nam_2011,ranjan_2012wicking,byon_2011} utilized Surface Evolver to generate the interfacial geometry \textit{a priori}. They then calculated the permeability and thermal resistance accordingly as inputs for the numerical simulations. These studies revealed that the increase in interface curvature enhances the evaporation flux by widening the thin-film region, whereas it deteriorates the liquid flow by reducing the wick permeability due to decreasing flow area \cite{nam_2011,ranjan_2012wicking}. The accuracy of different permeability models \cite{sobera_2006,sangani_1982,tamayol_2011,tamayol,srivastava_2010,zhang_2010,byon_2011,ranjan_2012wicking,xiao_2010} is experimentally and theoretically assessed on four different wick structures by Ravi \textit{et al.} \cite{ravi_2014}. They suggested a unified model that can predict the mass transfer with an error less than 18\% by combining the capillarity model of \cite{xiao_2010} and the permeability model of \cite{byon_2011}. For the better prediction of dryout heat flux, Zhu \textit{et al.} \cite{zhu_2016} developed a model by considering the variation of the meniscus shape with capillary pressure and local permeability. They also conducted validation studies by experimenting evaporation into the air atmosphere and reported the optimal wick geometry for the maximum dryout heat flux as functions of pillar diameter ($d$), height ($h$), and the pitch ($l$) as follows:  $d/h \sim 0.4-0.6$ and $l/d \sim 3.0$. 

In addition to the studies targeting the prediction of dryout heat load, there were other works \cite{ranjan_2009,ranjan_2011microscale,farokhnia_2016,ranjan_2012wicking,montazeri_2018,bongarala_2022} that focused on the estimation of interfacial heat flux by considering the thin-film evaporation or applying kinetic theory-based evaporation models. Ranjan \textit{et al.} \cite{ranjan_2009} investigated the wicking performance and effective thermal resistance for commonly used topologies as a function of a non-dimensional number that defines the characteristics of microstructures, liquid filling volume, and the contact angle. Farokhnia \textit{et al.} \cite{farokhnia_2016} conducted theoretical and experimental studies to systematically optimize interfacial heat flux for rectangular ribs, vertical circular, and square pillar configurations. They determined evaporative mass flux at the interface using the Hertz-Knudsen equation and suggested an optimum pitch-to-diameter ratio, $l/d \sim 1.8$, to maximize the heat flux by thin-film evaporation. Recently, Bongarala \textit{et al.} \cite{bongarala_2022} developed a non-dimensional metric (a figure of merit) to evaluate the evaporative heat flux for several wick structures and compared their predictions with those in previous works \cite{ranjan_2011microscale,ranjan_2012wicking,farokhnia_2016,montazeri_2018}. 

More comprehensive models were developed by coupling the fluid flow and thermal models to investigate the evaporation from cylindrical micropillars. Adera \textit{et al.} \cite{adera_2016} investigated the effect of porosity and pillar height for the optimum wick design to maximize the evaporation performance. They predicted the dryout heat flux based on a semi-analytical model similar to that of \cite{byon_2011} and estimated the interfacial thermal resistance using Schrage equation \cite{schrage}. The authors employed invariant permeability and thermal resistance values at the receding contact angle for the entire domain and neglected the curvature variation along the wicking direction. They also conducted verification experiments under a controlled environment sustaining steady evaporation into pure vapor and validated their model with 20\% accuracy. Wei \textit{et al.} \cite{wei_2018} carried out a parametric investigation to optimize the wick structure by conducting a wide range of experiments using varying micropillar geometries. In addition, the authors assessed their experimental results, specifically dryout heat flux and superheat, based on previous modeling efforts \cite{ravi_2014}. Somasundaram \textit{et al.} \cite{somasundaram_2018} compared the accuracy of existing permeability models and provided guidance for the optimal design to maximize dryout heat flux and minimize thermal resistance simultaneously. Recently, Vaarstra \textit{et al.} \cite{vaartstra_2019} developed a comprehensive model to apprehend arbitrary thermal load and nonuniform evaporation by extending the permeability model of \cite{zhu_2016}. The model accurately calculates the variation of the capillary pressure, heat transfer coefficient, and temperature distribution along the substrate.

Fundamentally, the heat transfer from a micropillar wick evaporator is a problem governed by various interfacial phenomena such as the capillarity induced liquid flow or thin-film evaporation intensifying near the contact lines. Thermocapillarity is also a well-known mechanism which induces a surface flow (commonly known as Marangoni flow) on non-isothermal interfaces. Its significant contribution to the convective transport, and the heat transfer thereof, was experimentally and theoretically demonstrated in different systems such as the evaporation of water droplets  \cite{ghasemi2010,kita2016,akdag2021}. However, the role of Marangoni convection in the micropillar wick evaporators has been overlooked mostly due to the complexity of the modeling for this problem. There were early modeling efforts to assess the effect of Marangoni flow in micropillar wicks \cite{ranjan_2011microscale,ranjan_2012wicking}. However, these attempts could not address the actual effect of thermocapillarity since the transport mechanisms were handled only at the cell-level, which prevented exploring its impact on the capillary flow along the substrate and the overall thermal performance thereof.

In the present study, we build a comprehensive model for simultaneous prediction of dryout heat flux and local temperature distribution for a micropillar wick evaporator. We extended the permeability and thermal resistance model developed in \cite{zhu_2016,vaartstra_2019} to capture the effect of thermocapillarity induced Marangoni convection in the cell-level model. The effect of Marangoni convection is then reflected on the device-level model in terms of permeability and effective thermal resistance for evaporation. The current model is validated with three controlled sets of distinct experiments to exhibit the model's capabilities in predicting dryout heat flux, heat transfer coefficient, and the effect of Marangoni flow.

\section{Methodology}
\label{sec:modeling}

\begin{figure}[h!]
\includegraphics[scale=0.85]{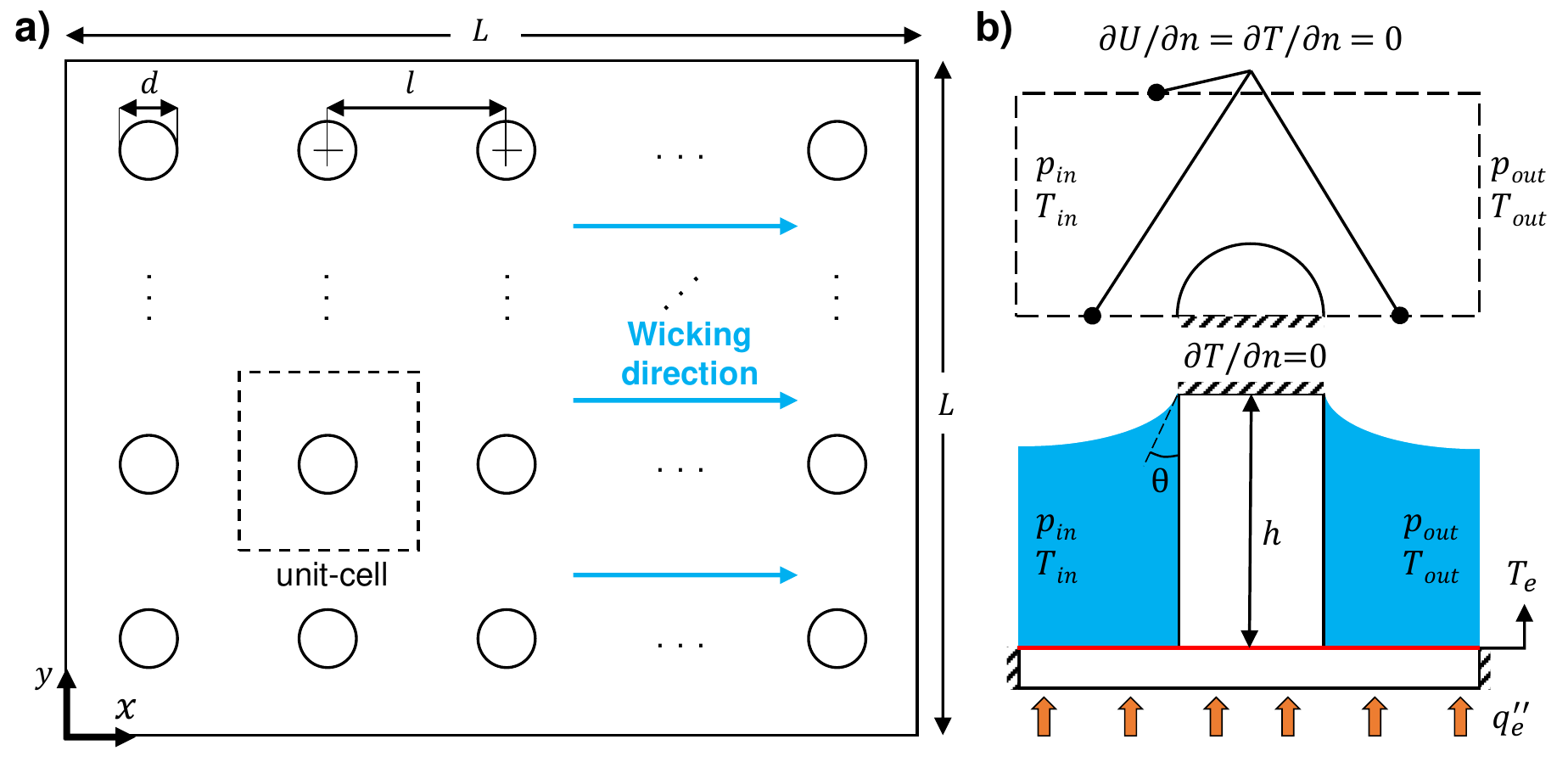}
\centering
\caption{\textbf{a)} Top view of the evaporator device. The pillar array dimensions are $d$, $l$, $h$, and $L$, where $d$ is diameter, $l$ is pitch, $h$ is height, $L$ is the total wicking length. \textbf{b)} Top and front view of the half unit-cell and boundary conditions.}
\label{fgr:close_Domain}
\end{figure}

The evaporation from regularly packed pillar arrays is investigated at the micro (cell-level) and macro (device-level) scales to obtain sufficiently accurate results with a reasonable computational cost. Device-level and cell-level computational domains are shown in \cref{fgr:close_Domain}. First, parametric studies are performed at the cell-level for a wide range of geometries and contact angles to obtain curvature-dependent permeability ($\kappa$) and effective heat transfer coefficient ($h_{\mathit{eff}}$) as a function of geometry and capillary pressure.  Then curvature-dependent local permeability and heat transfer coefficient values are utilized in the discretized device-level model, where all cells are linked to satisfy the conservation of mass, momentum, and energy throughout the substrate by coupling the energy and fluid transport problems. Dryout heat flux, local heat transfer coefficients, and temperature distribution on the substrate are acquired for an arbitrary thermal load applied at the bottom of the substrate. 

\subsection{Cell-level Model}

In the cell-level model, initially, 3-D meniscus shapes at various contact angles are generated (Sec.~\ref{sec:meniscus}). Then the generated 3-D geometries are employed as the computational domain for cell-level mass and energy transport in the presence of thermocapillarity (Sec.~\ref{sec:cell_cfd}).

\subsubsection{Modeling of Meniscus Shape}
\label{sec:meniscus}

Three-dimensional meniscus shape established between the pillars is a function of the local liquid pressure and vapor pressure.  Assuming constant vapor pressure along the evaporator and negligible variation of the curvature in a cell  ($l \ll L$), capillary pressure, \mbox{$p_c$}, is defined as the local liquid-vapor pressure difference \mbox{$p_c=p{_{v}}-p{_{l}}$}  that is calculated by the Young-Laplace (Y-L) equation:

\begin{equation} \label{eqn:Young-Laplace}
\nabla \cdot \mathbf{n} = p_c / \sigma 
\end{equation}

\noindent where $\sigma$  is surface tension, and $\mathbf{n}$ is the outward unit normal vector of the surface. The shape of interface is denoted as a geometric function of the surface height, $H$, and the unit outward normal of the interface is defined as follows:

\begin{equation} \label{eqn:unit-normal}
\mathbf{n} = \frac{(H_x, H_y, -1)}{(H_x^2+H_y^2+1)^{1/2}}
\end{equation}

\noindent where subscripts $x$ and $y$ represents the derivatives in the corresponding directions. After substituting \cref{eqn:unit-normal} into the Y-L equation, appropriate boundary conditions are needed to solve this nonlinear differential equation. Due to the symmetry of the meniscus within a unit-cell, half, quarter, or half-of-a-quarter of a unit-cell can be utilized as the computational domain. In every case, side boundaries possess symmetry boundary condition. Moreover, the three-phase contact line is fully pinned to the pillar top ($z$=0). 
Furthermore, the relation between the liquid-vapor pressure difference and contact angle is obtained by the force balance as follows: 

\begin{equation} \label{eqn:Pcap}
 p_c (\theta) = \frac{4 \sigma \cos\theta}{d \left[  \frac{4}{\pi}  \left( \frac{l}{d}   \right) ^{2} -1 \right]}
\end{equation}

\noindent where $\theta$ is the contact angle at the three-phase contact line, $l$ is the pitch, and $d$ is the diameter of the pillar. The balance between the upward surface tension forces on the pillar edge and the downward capillary pressure forces applied on the meniscus result in the static meniscus shape with solid-liquid contact angle $\theta$. The Y-L equation is solved to obtain the 3-D meniscus shape for a given capillary pressure and geometry. The generated 3-D liquid domain serves as the simulation domain for the cell-level model.

\begin{figure}[h!]
\includegraphics[scale=0.85]{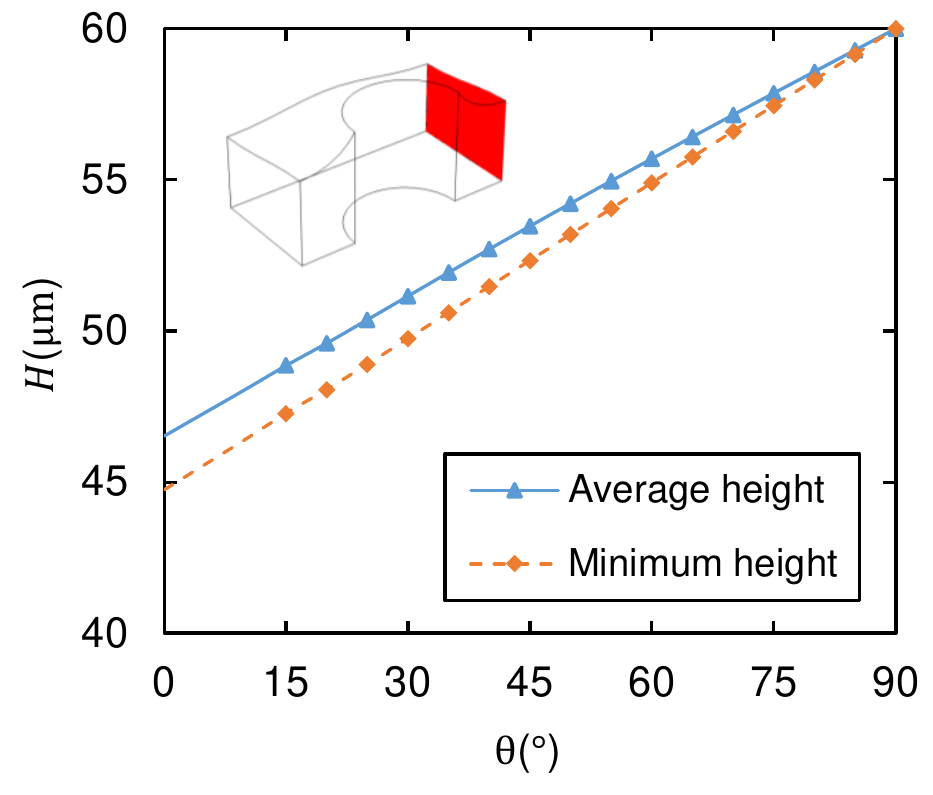}
\centering
\caption{ Variation of average and minimum meniscus height within a unit-cell ($d$=30$\, \rm \mu m$, $l$=60$\, \rm \mu m$, $h$=60$\, \rm \mu m$) as a function of contact angles. The red plane in the inset indicates the cross-sectional area at which average and minimum meniscus height are calculated.}
\label{fgr:meth_yl}
\end{figure}

A sweep solution for various contact angles is conducted between the flat ($\theta$=90$^\circ$) and fully stretched meniscus ($\theta$=$\theta_{rec}$) to capture the interface variation along the wicking direction. The variation of the average and minimum meniscus height as a function of contact angle is presented in the \cref{fgr:meth_yl}. The results exhibit linear-like behavior that makes solving at certain contact angles and interpolating the remaining angles possible. The accuracy of the Y-L equation solutions is ensured by comparing the 3-D meniscus shapes obtained from the simulations with those measured in the study of Adera \textit{et al.} \cite{adera_2016} in Supplementary Material.

\subsubsection{Modeling of Energy and Mass Transport}
\label{sec:cell_cfd}

Half of a unit-cell is considered as the computational domain for the multi-physics problem, where cell-level flow and thermal models are coupled in liquid and solid domains (see \cref{fgr:close_Domain}b). In the liquid domain, steady-state Navier-Stokes equations are utilized for the conservation of mass and linear momentum. Due to the dominating surface forces over the body forces, Bond number is low ($\rm{Bo}$ $=\Delta \rho  g l^{2}/\sigma \sim \mathcal{O} (10^{-4})$); therefore, gravity is omitted in the modeling. To account for the thermocapillary driven convective energy transport, a full steady-state energy equation (with convective terms) is utilized. The flow is laminar ($\rm Re$ $=\rho u  l/ \mu \sim \mathcal{O} \left( 10^{1} \right) $), and the viscous dissipation term is not included in the energy equation because of the low liquid velocities. The computational model utilizes temperature-dependent properties throughout the domain, which is specifically crucial to include the effect of viscosity variation. Governing equations are summarized as follows:

\begin{subequations}
\begin{equation} \label{eqn:vector-cont}
\nabla\cdot(\rho\mathbf{u}) = 0
\end{equation}
\begin{equation} \label{eqn:vector-mom}
\rho(\mathbf{u}\cdot\nabla)\mathbf{u} = -\nabla p + \nabla \cdot \bar{\bar{\tau}}
\end{equation}
\begin{equation} \label{eqn:vector-en}
\rho c_p\mathbf{u}\cdot\nabla T = \nabla \cdot (k\nabla T)
\end{equation}
\end{subequations}

\noindent where $\rho$, $c_p$, and $k$ are density, specific heat, and thermal conductivity of the liquid, respectively; $\mathbf{u}$, $p$, and $T$  are velocity, pressure, and temperature of the liquid. $\bar{\bar{\tau}}$ is the deviatoric stress tensor defined as \mbox{$\mu ( \partial u_{i} / \partial x_{j} + \partial u_{j} / \partial x_{i} )$}, where $\mu$ is the liquid viscosity. In the solid domain, on the other hand, governing equations reduce to the steady-state heat equation: $\nabla^{2}T=0$. The values of the temperature dependent properties are taken from the material library of COMSOL Multi-physics software \cite{comsol}.

The next task is to determine the flow and thermal boundary conditions for the computational domain, \textit{i.e.} the half unit-cell. For both flow and thermal problems, symmetry conditions ($\del_{n} \mathbf{u}= 0$, $\del_n T = 0$) are utilized at the boundaries parallel to the wick (capillary) flow. At the boundaries perpendicular to the capillary flow, inlet and outlet conditions are determined based on periodicity. For the flow, periodic hydrodynamic condition ($\mathbf{u_{in}}=\mathbf{u_{out}}$) with the pressure difference between adjacent cells ($p_{in}-p_{out}=\Delta p_c$) is applied to ensure fully developed liquid flow. For the thermal modeling, application of periodic boundary condition ($T_i=T_{out}$) implies no convective energy transport across cells and this approximation is justified by the negligible temperature variation between adjacent cells.
On the solid-liquid contacts no-slip boundary condition ($\mathbf{u}=0$) is utilized. Heat transfer from the pillar top \textit{via} natural convection is negligible compared to the evaporative heat transfer from the interface. A uniform heating ($q={q}''_{e}$) is applied at the bottom surface of the substrate in a cell.  


Mass, force, and energy balances determine the boundary conditions at the liquid-vapor interface. The mass conservation at the interface requires the balance of outflow and evaporating mass flow as follows:

\begin{equation} \label{eqn:mass_balance}
\mathbf{u} \cdot \mathbf{n}=\dot{m}''_{evap}/{\rho}
\end{equation}

\noindent where $\dot{m}''_{evap}$ is the evaporative mass flux. It should be noted that \textit{a posteriori} analysis shows evaporation-induced normal liquid velocity is dominated by the thermocapillarity-induced tangential velocity. Consequently, normal liquid flow can be negated in the modeling. 

While the normal force balance is established \textit{via} Young-Laplace equation, tangential force balance requires the consideration of thermocapillarity. Marangoni (thermocapillary) flow is driven by the surface tension variation on the liquid-vapor interface originating from the temperature gradients. The impact of thermocapillary flow can be measured with the Marangoni number, which is defined as: 

\begin{equation} \label{eqn:ma_1}
\textrm{Ma} = \frac{\partial{\sigma}}{\partial{T}} {\frac{L_t \Delta T_{int}}{\mu\alpha}}
\end{equation}

\noindent where $L_t$ is the length scale of the domain (the tangential distance on the meniscus between the pillar edge and a corner of the unit-cell), $\Delta T_{int}$ is the maximum temperature difference at the interface, and $\alpha$ is thermal diffusivity of the liquid. Accordingly, in the case of high superheat and/or large pitch, thermocapillary flow can trigger effective internal convection in a cell. Tangential force balance is applied between the thermocapillary and the shear force associated with the liquid at the liquid-vapor interface as follows:

\begin{equation} \label{eqn:ma}
-\mathbf{n} \cdot {\bar{\bar{\tau}}} \cdot \mathbf{t} = \nabla\sigma \cdot \mathbf{t}
\end{equation}

\noindent where $\mathbf{t}$ is the unit tangential vector. In the tangentail force balance shear stress induced by the gas phase is neglected due to
lower viscosity of the vapor. 

Interfacial energy balance is established by several physical mechanisms. The primary mechanism is the evaporation, which is associated with the breaking of physical bonds between liquid molecules. Conduction to the gas phase and radiation to the surroundings are also present. Yet, the current modeling does not include the gas phase and the enclosure surrounding it. Consequently, estimation of heat transfer for these mechanisms cannot be made precisely. Moreover, conduction across the Knudsen layer is commonly insignificant in the absence of non-condensable gases \cite{vaartstra2020capillary}. Accordingly, evaporative heat transfer dominates the others in the problem of interest. Therefore, the energy transported to the interface is assumed to be utilized by the evaporation solely: 

\begin{equation} \label{eqn:energy_balance}
\mathbf{n} \cdot (-k \nabla T) = \dot{m}''_{evap}h_{\mathit{fg}}
\end{equation}

\noindent $h_{\mathit{fg}}$ is the latent heat of vaporization. Since the current problem considers the evaporation of liquid into its own vapor, kinetic theory can be utilized for the estimation of evaporative mass flux. Utilization of Hertz-Knudsen equation is not practical due to the need for two experimental (evaporation and condensation) coefficients. Schrage's \cite{schrage} approach incorporating the effect of drift velocity has the advantage of the utilization of single experimental coefficient, commonly referred as to mass accommodation coefficient (MAC). When the evaporation rate is not excessive, Schrage's original expression \cite{schrage} reduces to an approximate form as reviewed in \cite{carey}, which is shown below:

\begin{equation} \label{eqn:m_evaporative}
\dot{m}''_{evap} = \frac{2 \hat{\sigma}}{2 - \hat{\sigma}} \left( \frac{M}{2 \pi R_u} \right)^{1/2}
\left( \frac{p_{sat}|_{T_{lv}}}{T_{lv}^{1/2}} - \frac{p_v}{T_v^{1/2}} \right)
\end{equation}

\noindent  where \mbox{$\hat{\sigma}$}, $M$, $R_u$, $T_{lv}$, $p_{sat}|_{T_{lv}}$, $T_v$, and $p_v$ are the MAC, molar mass of the liquid, universal gas constant, interface temperature, saturation pressure at the interface temperature, vapor temperature, and vapor pressure, respectively. A possible drawback of Schrage expression for the problem of interest could be its limitation for the use in curved interfaces since it was derived for a flat liquid-vapor interface. As suggested by Wayner and his coworkers \cite{wayner1976,wayner1991}, based on the assumption of small superheat ($T_{lv}^{1/2}=T_v^{1/2}$), Schrage's approximate expression can be converted to a form, which incorporates Clapeyron (\textit{i.e.} superheat) and Kelvin (\textit{i.e.} curvature) effects, simultaneously (see Ref. \cite{akkus2017} for the details of the derivation). However, the validity of this form was reported to be restricted with the superheat value of nearly 5$\,$K  \cite{wang2007characteristics}. In the problem of interest, superheat values are substantially higher than 5$\,$K; therefore, Schrage's expression (\cref{eqn:m_evaporative}) is utilized to estimate evaporation rate. Yet, we assess the error associated with the omission of the curvature effect for varying degrees of curvature in the Supplementary Material and demonstrate that the deviation is negligible for the micropillar evaporators modeled in the current study.

Applying aforementioned boundary conditions together with the interfacial mass, force, and energy balance considerations, governing equations are solved simultaneously using the FEM solver of COMSOL Multiphysics software. Internal grid generator of the software is utilized to create the solution mesh and the mesh independence study is presented in Supplementary Material.

\begin{figure}[b!]
\includegraphics[scale=0.85]{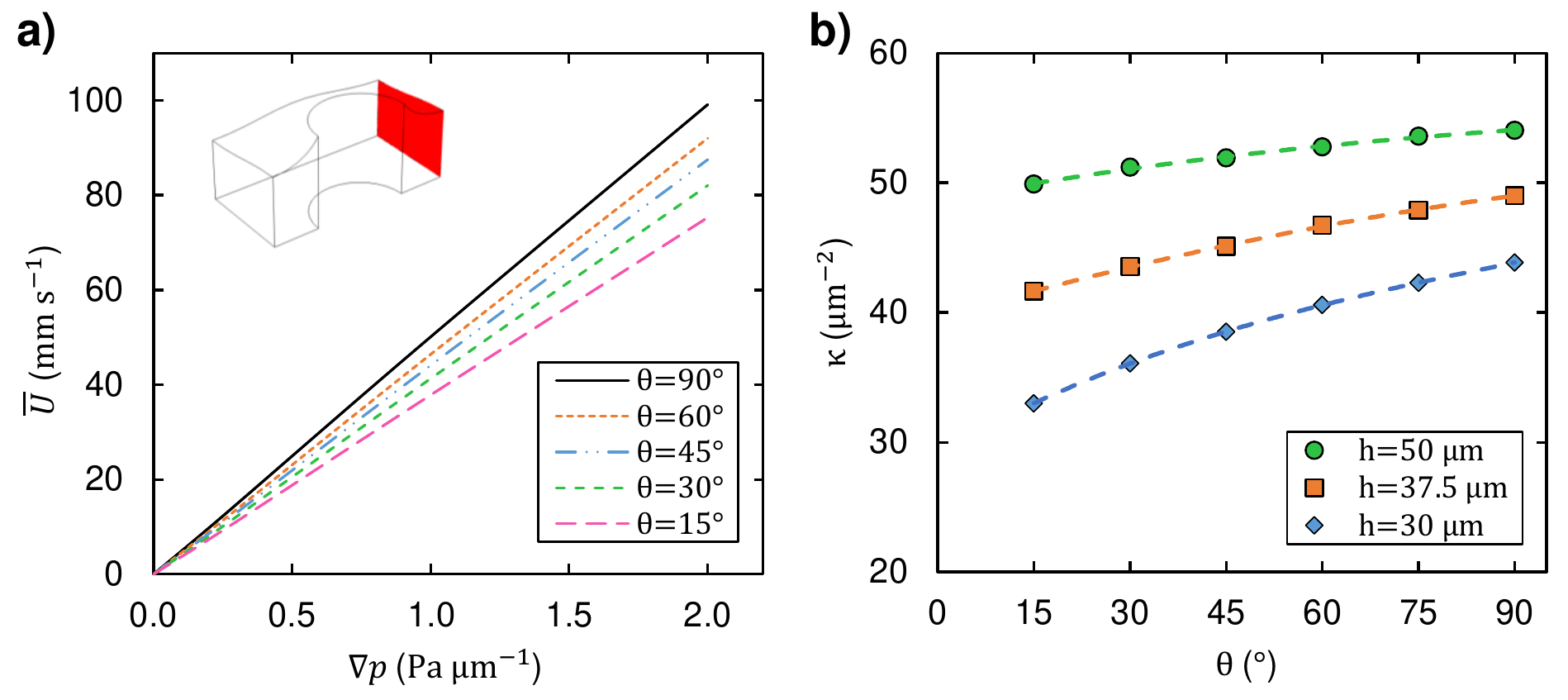}
\centering
\caption{\textbf{a)} The average velocity of water as a function pressure gradient at different contact angles for a pillar structure with following dimensions: $d$=30$\, \rm \mu m$, $l$=60$\, \rm \mu m$, $h$=30$\, \rm \mu m$. The red plane in the inset indicates the cross-sectional area ($A_c$) at which average velocity and permeability of the wick structure at the corresponding contact angle is calculated. \textbf{b)} Permeability as a function contact angle for three pillar geometries ($d$=30$\, \rm \mu m$, $l$=60$\, \rm \mu m$) with different heights. Dashed lines are the third-order polynomial fits to the data.}
\label{fgr:vel_and_kappa}
\end{figure}

Based on the resultant flow, the liquid permeability of the 3-D pillar structure in the wicking direction ($\kappa$) is obtained since it serves as an input for the device-level model. The resultant flow rate is utilized to back-calculate the permeability from Darcy's law:

\begin{equation} \label{eqn:U_ave}
\overline{U} = \frac{1}{A_c} \iint u \,dy\,dz = - \frac{\kappa}{\mu} \nabla p
\end{equation}

\noindent where $\overline{U}$  is the average velocity, ${u}$ is the velocity component in the $x$-direction, $A_c$ is the cross-sectional area at the outlet (see the inset in \cref{fgr:vel_and_kappa}a), $\nabla p$ is the pressure gradient, which is the pressure difference over a pitch of the pillar array, \textit{i.e.} $\nabla p = \Delta p_c / l$.
For simplicity, parametric simulations are conducted with a reference temperature of the liquid equal to the vapor temperature. However, in the device-level simulations, viscosity is modified as a function of the average liquid temperature that increases due to the applied heat flux. For a set of given pillar array dimensions, the capillary pressure defines the curvature of the interface (or the contact angle) and permeability ($\kappa$) as a function of the local pressure. Solving average velocities for various $\theta$ provides the relation between local permeability and capillary pressure.  The effect of pressure gradient over the average velocity is presented in the \cref{fgr:vel_and_kappa}a at different contact angles in a unit cell, and $\overline{U}$ is linearly varying for a wide range of applied pressure gradient. The permeability values at different contact angles (or capillary pressure) are presented in \cref{fgr:vel_and_kappa}b for three exemplary wicking structures with varying heights. As the pillar height increases, the permeability increases, and the effect of the contact angle on the permeability diminishes. To calculate the permeability of the wick structure at different capillary pressure (or contact angles) in the device-level model, a third-order polynomial fits, as shown in \cref{fgr:vel_and_kappa}b, are utilized. 

\begin{figure}[h!]
\includegraphics[scale=0.85]{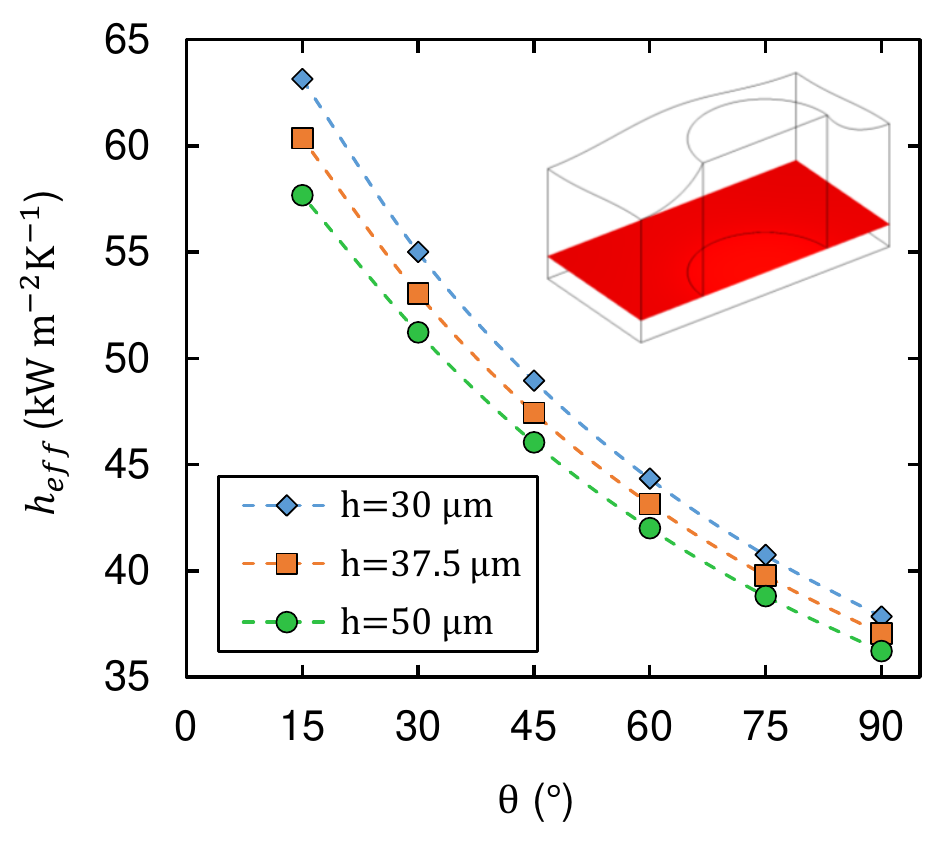}
\centering
\caption{Effective heat transfer coefficient as a function of contact angle for three pillar geometries ($d$=30$\, \rm \mu m$, $l$=60$\, \rm \mu m$) with different heights. The red plane in the inset indicates the liquid-substrate contact surface, where average temperature, $T_e$, is calculated to obtain $h_{\mathit{eff}}$ for the wick structure. Dashed lines are the third-order polynomial fits to the data.}
\label{fgr:meth_heff}
\end{figure}

Based on the resultant temperature field, the effective heat transfer coefficient is obtained since it serves as an input for the device-level model. The effective heat transfer coefficient is defined in terms of the evaporator heat flux, $q''_{e}$, and the average evaporator superheat, $\Delta T$, as follows: $h_{\mathit{eff}} = q''_{e} / \Delta T$, where the average evaporator superheat is the difference between average evaporator temperature, $T_{e}$, calculated at the liquid-substrate contact surface (see the inset in \cref{fgr:meth_heff}) and the vapor temperature, $T_v$: $\Delta T=T_{e} - T_v$.   The effective heat transfer coefficient values at different contact angles (or capillary pressure) are presented in \cref{fgr:meth_heff} for three wicking structures with varying heights. The thin-film area extends as the contact angle decreases due to the improved curvature; therefore, in \cref{fgr:meth_heff}, a substantial enhancement in $h_{\mathit{eff}}$ is observed as contact angles diminish. As the pillar height increases, the effective heat transfer coefficient decreases due the additional conduction resistance. To calculate the thermal resistance at different capillary pressure (or contact angles) in the device-level model, a third-order polynomial fits, as shown in \cref{fgr:meth_heff}, are utilized.

Details regarding the calculation permeability ($\kappa$) and effective heat transfer coefficients ($h_{\mathit{eff}}$) have been provided. Before skipping to the device-level modeling, $\kappa$ and $h_{\mathit{eff}}$ at various heat fluxes are stored to create lookup tables for a given geometry. Parametric simulations are performed at different contact angles and heat fluxes for a wide range of pillar array dimensions to observe the impact of the thermocapillary flow in the device-level model.

\subsection{Device-level Model}
\begin{figure}[b!]
\includegraphics[scale=0.85]{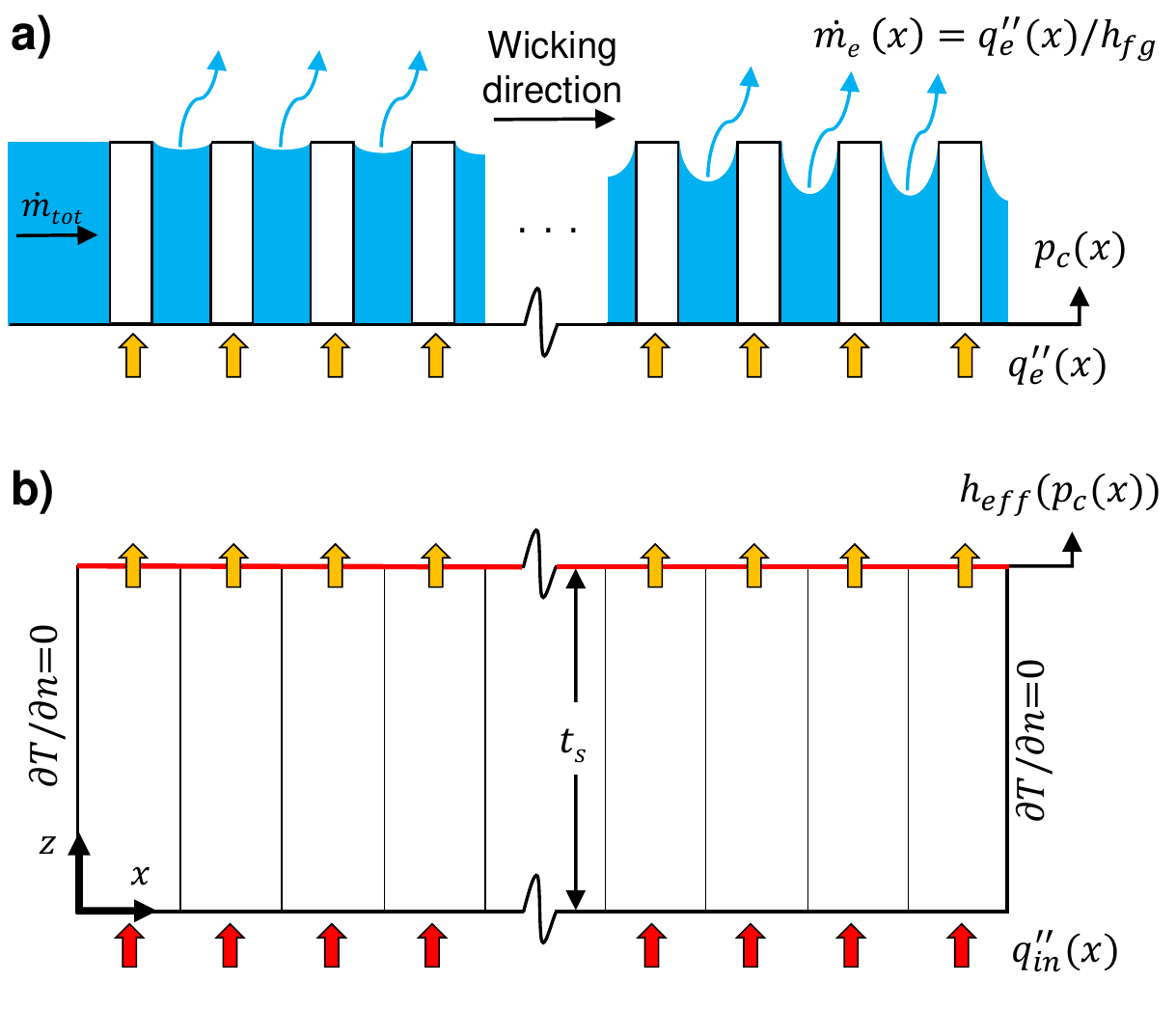}
\centering
\caption{Representations of the device-level \textbf{a)} flow and \textbf{b)} conduction domains. }
\label{fgr:close_Device}
\end{figure}

In the device-level model, evaporation is compensated by replenishing flow, which is driven by the variance of the capillary pressure along the wicking direction. The liquid replenishing is maintained until the contact angle of the meniscus in a unit-cell reaches the receding contact angle, which is herein denominated as the receding onset, and the corresponding heat flux is called dryout heat flux for a given pillar array dimensions. In the device-level simulations, local capillary pressures, average fluid temperature, and temperature distributions on the substrate are calculated. The device-level model is established based on the flow (see \cref{fgr:close_Device}a) and conduction (see \cref{fgr:close_Device}b) domains, which are discretized in accordance with the pitch of the unit-cells utilized in the cell-level modeling. 

First, the flow domain is considered to obtain the capillary pressure variation along the wicking direction for a predefined heat flux. The contact angle at the inlet is assumed as 90$\degree$ ($p_c$=0) since the wicking structure is in contact with a liquid reservoir. 
Since the applied heat flux is removed solely by the evaporation, the corresponding total liquid mass flow \mbox{($\dot{m}''_{tot}=q''_{in} l L /h_{\mathit{fg}}$)} is calculated as the inlet boundary condition. The capillary pressure and mass flux at the inlet are sufficient to acquire capillary pressure distribution over the entire domain since the pressure drop depends on permeability ($\kappa$) and mass flow rate ($\rho A_c \overline{U}$) according to Darcy's law.

\begin{figure}[b!]
\includegraphics[scale=0.85]{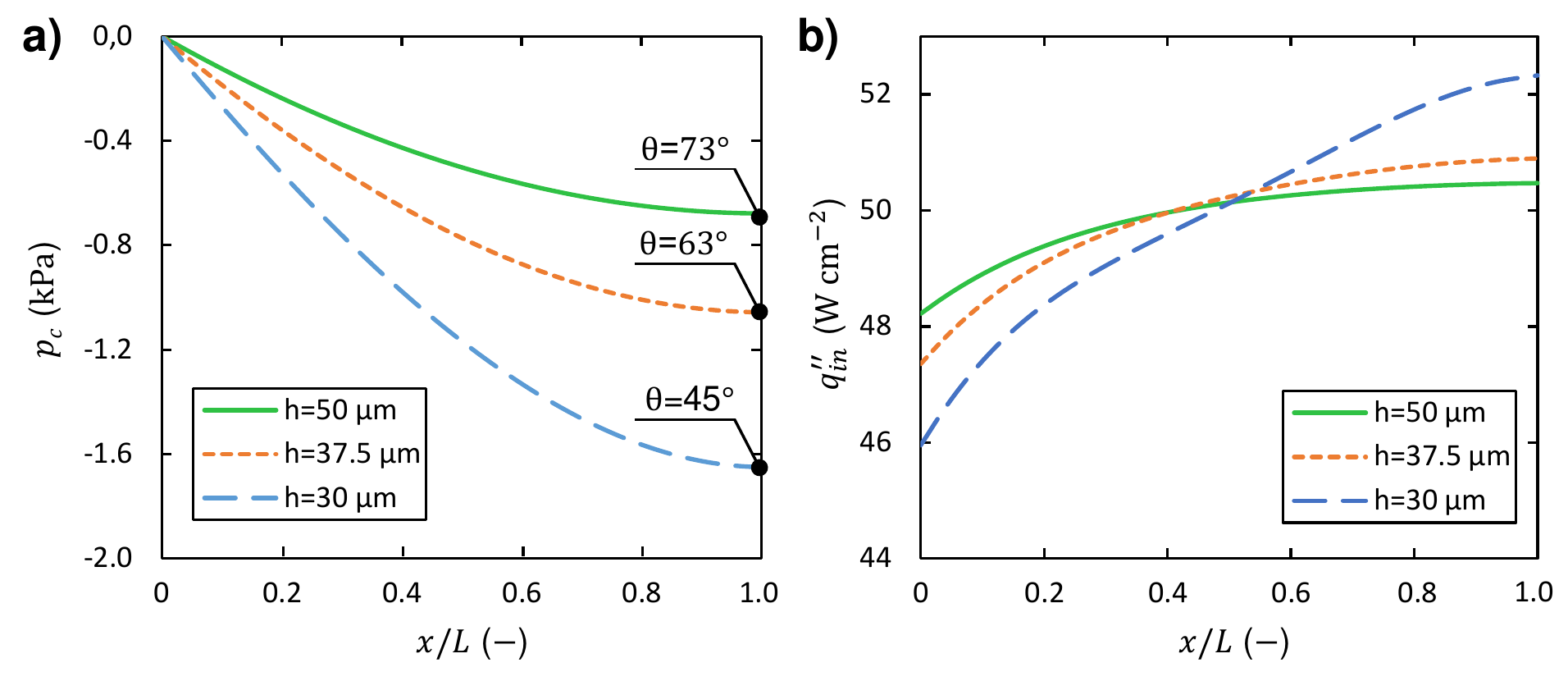}
\centering
\caption{Distributions of the \textbf{a)} capillary pressure and \textbf{b)} evaporator flux along the wicking direction of three wick structures ($d$=30$\, \rm \mu m$, $l$=60$\, \rm \mu m$) with different heights as the result of device-level model. Uniform heat flux of $q''_ {in}$=50$\, \rm W \, cm^{-2}$ is applied at the bottom of the base substrate ($t_s$=600$\, \rm \mu m$, $L$=5$\, \rm mm$).}
\label{fgr:device_x}
\end{figure}

After the calculation of \mbox{$p_c (x)$} along the flow domain, the spatial distribution of {\mbox{$h_{\mathit{eff}}(p_c)$}} is obtained and going to be utilized in the device-level conduction domain, where a heat input, \mbox{$q''_{in} (x)$}, is applied at the bottom surface of the substrate with a base thickness of $t_s$ and heat transfer from the side walls of the substrate is negated. In the first iteration, \mbox{$p_c (x)$} was calculated by neglecting the axial conduction in the substrate. However, the applied heat flux at the bottom of the substrate, \mbox{$q''_{in} (x)$}, results in nonuniform evaporation through the axial conduction due to the spatial variation of $h_{\mathit{eff}}$. Consequently, the heat input applied to the flow domain, \textit{i.e.} the evaporator heat flux, \mbox{$q''_{e} (x)$}, acquires a new distribution. Therefore, device-level flow domain needs to be re-visited to update the flow solution based on the updated distribution of \mbox{$q''_{e} (x)$} from solution of device-level conduction problem. Conduction and flow problems are solved iteratively by an outer loop till the convergence. The computational scheme of the complete model is presented in Supplementary Material. Resultant distributions of capillary pressure and evaporator heat flux along the substrate at the end of the iterative solution are provided for three exemplary wicking structures with varying heights in \cref{fgr:device_x}a and b, respectively.
Wick structures with taller pillars have higher contact angle at the end of the wick ($x$=5$\,$mm) due to the decreased need for capillary pumping associated with their higher permeability. Near the end of the wick, variation of capillary pressure becomes less pronounced due to the decreased amount of liquid flow. On the other hand, the resultant variation of $h_{\mathit{eff}}$ along the wicking direction enables nonuniform heating, and nonuniform evaporation from the interface thereof, as shown in \cref{fgr:device_x}b. This effect is as a result of the conduction heat transfer through the substrate. Nonuniform evaporation is more pronounced for shorter pillars because of the larger variation of the film thickness along the wick.

\section{Results and Discussion}
\label{sec:results}

The current model is validated against three distinct sets of experiments conducted in previous studies \cite{zhu_2016,adera_2016,wei_2018}, where all substrates were made of silicon with a wide variety of pillar dimensions, and the working liquid was deionized water. Varying heat inputs were applied under different ambient conditions. The substrates with similar pillar dimensions led to near predictions for the dryout heat flux or superheat values. Accordingly, selected cases from these studies (see \cref{tab:case1table}) are included in the validation of the model for brevity. Pillar dimensions were sufficiently small for all experiments to prevent the incipience of boiling.

\begin{table}[b!]
\caption{Details of the micropillar wick evaporators and experimental conditions in previous experiments \cite{zhu_2016,adera_2016,wei_2018}, which are simulated by the proposed model.}
\centering
\vspace*{5pt}
\begin{tabular}{lccc c c c cc}
\hline\\[-10pt]
 & \multicolumn{3}{c}{\textbf{Dimensions ($\mu m$)}} & & & & \\[2pt]
 \cline{2-4}  \\[-10pt]
\textbf{Device ID} & \textbf{$d$} &  \textbf{$l$} & \textbf{$h$} & $\theta$ (\degree) &{$T_{sat}$(\degree C)} & $q''_{in}$(W cm$^{-2}$) & Ma   \\[5pt] 
\hline\\[-10pt]
Device-A1	\cite{zhu_2016}	&	7	&	20	&	20	& 90$\, \rightarrow$15 &	100$^{*}$	&	0--47	&		-	\\[5pt] 
Device-A3	\cite{zhu_2016}	&	6	&	50	&	19	& 90$\, \rightarrow$15 &	100$^{*}$	&	0--27	&	-	\\[5pt] 
Device-1	\cite{adera_2016}	&	5	&	12	&	82 & 90$\, \rightarrow$70	&	24	&	0--46	&		36--30	\\[5pt] 
Device-5	\cite{adera_2016}	&	12	&	20	&	90	& 90$\, \rightarrow$70 &	24	&	0--45	&	74--62	\\[5pt] 
Sample-1	\cite{wei_2018}	&	23.4	&	41.0 	&	39.5 & 90$\, \rightarrow$10	&	25	&	0--28	&		73--40	\\[5pt] 
Sample-2	\cite{wei_2018}	&	25.9	&	41.0	&	39.5 & 90$\, \rightarrow$10	&	25	&	0--22	&		49--32	\\[5pt] 
Sample-3	\cite{wei_2018}	&	28.2	&	41.0	&	39.5 & 90$\, \rightarrow$10	&	25	&	0--23	&		36--23	\\[5pt] 
Sample-4	\cite{wei_2018}	&	22.1	&	41.0	&	39.5 & 90$\, \rightarrow$10	&	25	&	0--23	&		77--51	\\[5pt] 
Sample-5	\cite{wei_2018}	&	23.4	&	36.0	&	39.5 & 90$\, \rightarrow$10	&	25	&	0--22	&		32--20	\\[5pt] 
Sample-6	\cite{wei_2018}	&	23.4	&	34.0	&	39.5 & 90$\, \rightarrow$10	&	25	&	0--20	&		22--14	\\[5pt] 
Sample-7	\cite{wei_2018}	&	23.4	&	44.0	&	39.5 & 90$\, \rightarrow$10	&	25	&	0--25	&		97--64	\\[5pt] 
Sample-8	\cite{wei_2018}	&	23.2	&	41.0	&	79.7 & 90$\, \rightarrow$10	&	25	&	0--60	&		293--215	\\[5pt] 
\hline
\multicolumn{8}{l} {$^{*}$ Evaporation into the air environment at 1 atm.}
\end{tabular}
\label{tab:case1table}
\end{table}

\begin{figure}[h!]
\includegraphics[scale=0.85]{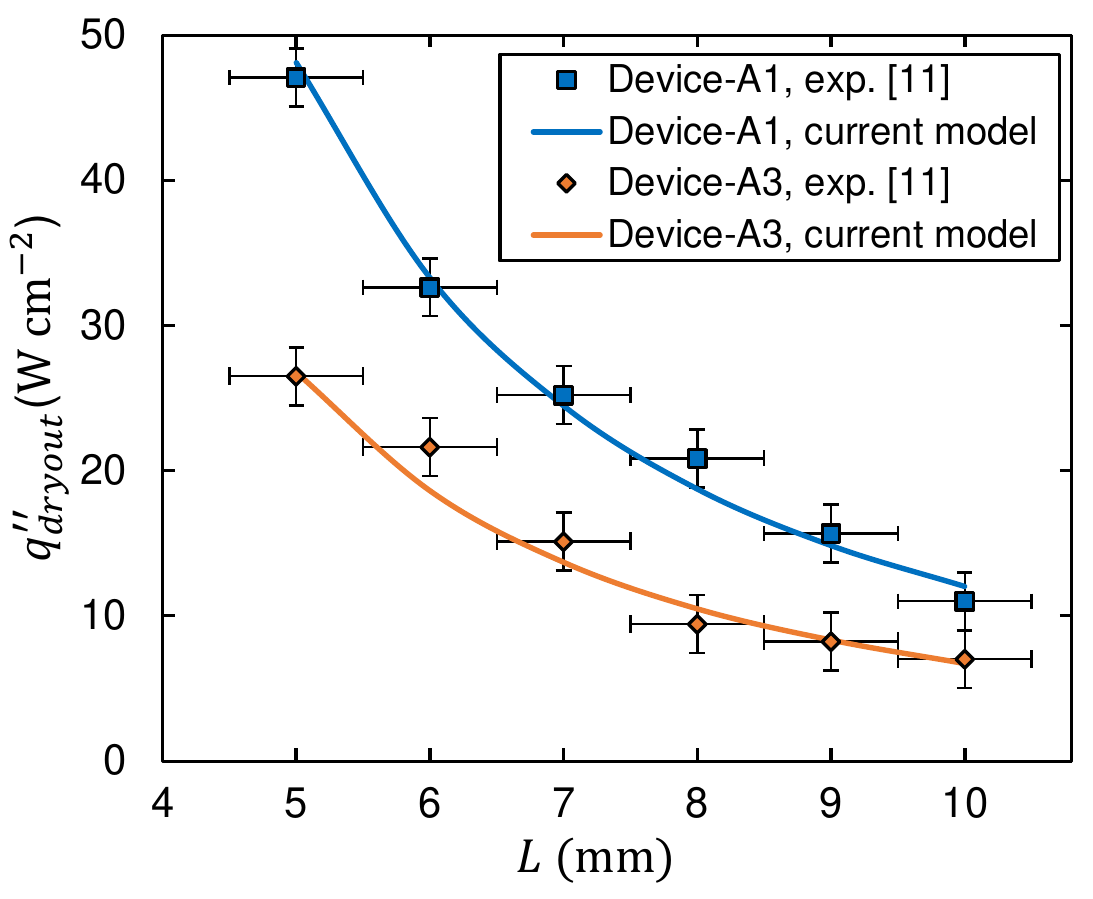}
\centering
\caption{Comparison of the dryout heat flux predictions of the proposed model and the experimental results of Zhu \textit{et al.} \cite{zhu_2016}.}
\label{fgr:result_zhu}
\end{figure}

In the first experimental work \cite{zhu_2016}, simulated by the proposed model, evaporation took place into the air from two different samples (Device-A1 and Device-A3 in \cref{tab:case1table}). While the pillar diameter and height were close, pitches of the pillars were different ($20 \rm \,\mu m$\,vs.\,$50 \rm \, \mu m$) in these substrates. Accordingly, the wicking ability of the substrates differs substantially. The substrate with a denser pillar forest could withstand a higher heat load without drying for the same liquid wicking length in the experiments. The predictions of the proposed model excellently match with the experimental results as shown in \cref{fgr:result_zhu}, where maximum wicking lengths at the onset of dryout are presented for different heat fluxes. While this comparison reflects the success of the proposed model in terms of liquid transport, the sensitivity of the model to the thermal effects could not be assessed since no temperature data were available.

\begin{figure}[h!]
\includegraphics[scale=0.85]{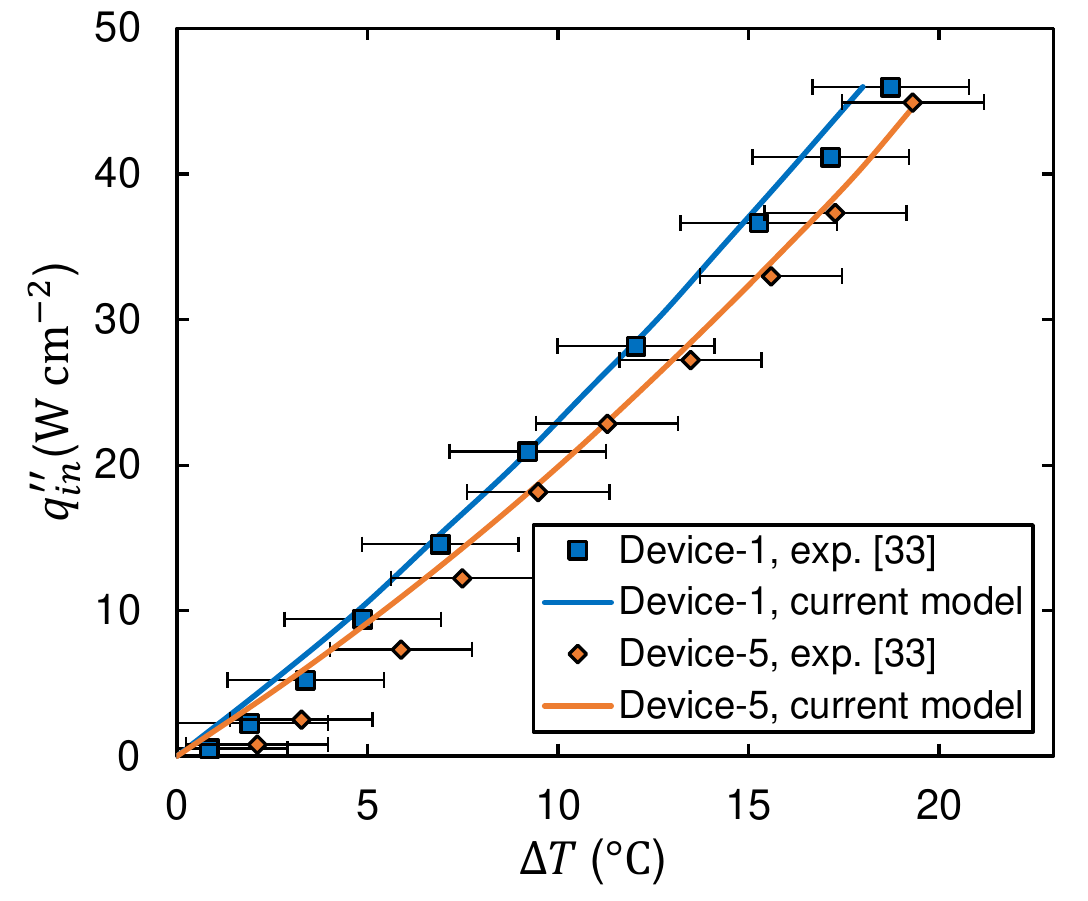}
\centering
\caption{Comparison of the average evaporator superheat predictions of the proposed model and the experimental results of Adera \textit{et al.} \cite{adera_2016}. For both devices, MAC value of 0.06 is utilized in the estimation of evaporation rates.}
\label{fgr:result_adera}
\end{figure}

Secondly, the experimental work of Adera \textit{et al.} \cite{adera_2016} is utilized to verify the thermal capabilities of the proposed model. In their experiments, the evaporation took place in a controlled environment with pure vapor, and the wicks were in contact with the liquid at the four sides of the substrate. Accordingly,  bi-directional liquid flow is considered in our modeling to represent the physics accurately (see Supplementary Material for the details of bi-directional flow modeling). A relatively high receding contact angle ($\theta_{rec}$=70\degree) was observed in the experiments, which was attributed to the presence of polymer remaining on the pillars. The substrate temperature was measured at different heat fluxes by gradually increasing the applied heat input, and the average temperature at the evaporator was estimated through a one-dimensional thermal resistance model. These average evaporator temperature estimations of Adera \textit{et al.} \cite{adera_2016} are successfully predicted by the proposed model as shown in \cref{fgr:result_adera}, where the average temperature is represented by the average evaporator superheat.

It should be noted that evaporator temperature is primarily dictated by the rate of evaporation, which determines the extent of the evaporative cooling. Therefore, calculation of the evaporation rate \textit{via} Schrage's approximate expression (\cref{eqn:m_evaporative}) has a substantial effect on the predicted temperatures in the current model. However, Schrage's expression includes a parameter, \mbox{$\hat{\sigma}$}, the mass accommodation coefficient or MAC in short, which does not possess a common universal value. In fact, it is a system-dependent parameter \cite{somasundaram_2018}, which varies with the type of the liquid, system cleanliness, etc. From the kinetic theory point of view, a departure from the equilibrium alters the macroscopic vapor (drift) velocity and the thermophysical properties in the vapor. These effects influence the value of MAC utilized in the estimation of evaporation since evaporation is inherently a phenomenon shifting the equilibrium of the state. Further discussion on this subject is out of the scope of this study. Past studies focusing on MAC \cite{akkus2021drifting}, or evaporation/condensation coefficients \cite{persad2016} offer a broader perspective.

The proposed strategy in the current work was to determine a MAC value for each device and utilize this value for the experiments conducted on the same device. A system-specific MAC is obtained with this approach, which enables accounting for the uncontrollable parameters associated with the experimentation. We, however, refrain from assigning different MAC values to each experiment since it would not be practical when modeling is performed for design purposes, and it might break the connection between MAC and the system by turning MAC into a fudge factor serving as a fitting parameter between each simulation and experiment.   

The presence of a small amount of contamination on the water interface is known to hinder the induction of a Marangoni flow  \cite{hu2005}. Therefore, the presence of polymer remaining on the pillars fabricated for the study of Adera \textit{et al.} \cite{adera_2016} makes those experiments inappropriate for the investigation of the effect of thermocapillary flow on the evaporation from micropillar evaporators. On the other hand, the study of Wei \textit{et al.} \cite{wei_2018} provides a suitable base for the investigation of thermocapillary flow by providing a set of evaporation experiments into pure vapor environment across a wide range of Ma numbers (see \cref{tab:case1table}). Moreover, the observation of the low receding contact angle of water ($\theta_{rec}$=10\degree) suggests contamination-free pillars. Therefore, the current model is applied to simulate the experiments of Wei \textit{et al.} \cite{wei_2018}. 
Three different scenarios are modeled to assess the individual effects of temperature variation and Marangoni flow: i) isothermal model, ii) non-isothermal model, and iii) Marangoni model. In the isothermal model, the liquid temperature is set to the vapor temperature, and the liquid temperature increase associated with the applied heat load is neglected. But this simplification directly affects the thermophysical properties, especially the liquid viscosity, which is appreciably sensitive to the temperature. In the non-isothermal model, on the other hand, the liquid temperature increase is obtained from the cell-level thermal model, and the modeling is performed based on temperature-dependent thermophysical properties. In the Marangoni model, thermocapillary convection is also included such that all active transport mechanisms are taken into consideration in the model.

\begin{figure}[b!]
\includegraphics[scale=0.75]{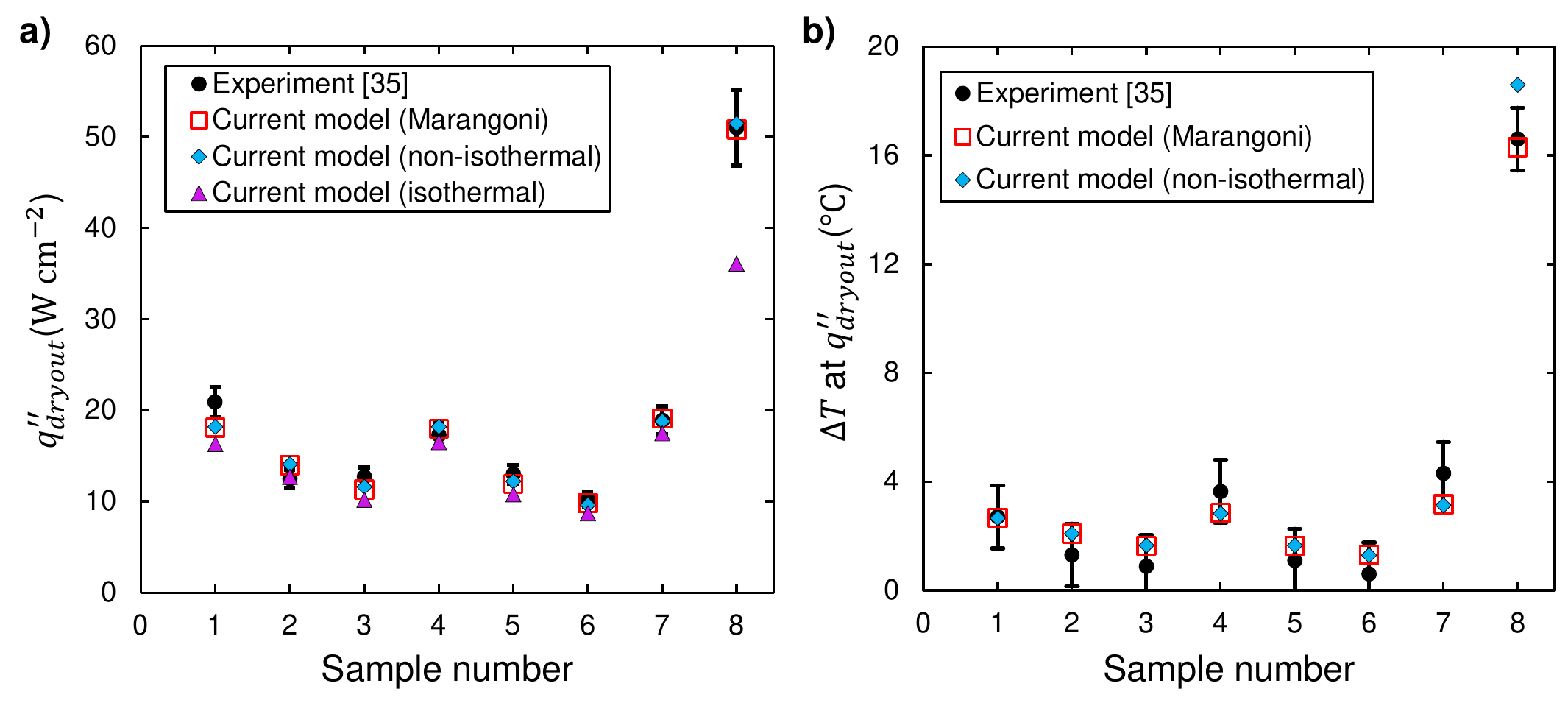}
\centering
\caption{Comparison of the \textbf{a)} dryout heat flux and \textbf{b)} average evaporator superheat predictions of the proposed model and the experimental results of Wei \textit{et al.} \cite{wei_2018}. Different MAC values between 0.1 and 0.4 are assigned to each sample and the assigned value is utilized for all experiments conducted on the corresponding sample in the estimation of evaporation rates.}
\label{fgr:result_wei}
\end{figure}

\begin{figure}[t!]
\includegraphics[scale=0.7]{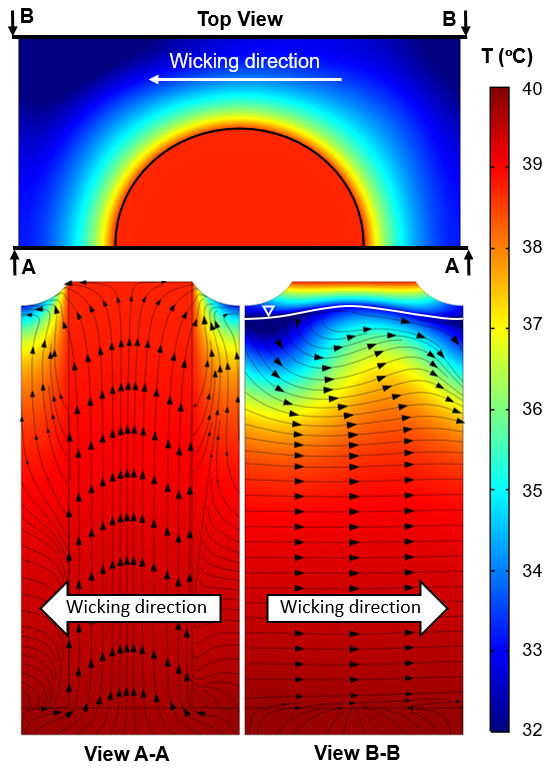}
\centering
\caption{Temperature field and superimposed total energy flux streamlines on the interface (top view), pillar center plane (view A-A), and symmetry center plane between adjacent unit-cells (view B-B). In view B-B, the liquid-vapor interface is shown by white solid line. Results are given for Sample-8 experimented in Ref. \cite{wei_2018} at a heat load of $q''_ {in}$=50$\,  \rm W \, cm^{-2}$ for a unit-cell with the contact angle of $\theta$=30$ \degree$.}
\label{fgr:Ma_temp}
\end{figure}

The first comparison is made for the dryout heat fluxes as shown in \cref{fgr:result_wei}a.
Since the temperature rise is relatively restricted for the first seven samples, predictions of isothermal and non-isothermal models are close. However, for Sample-8, which has the highest heat input, the isothermal model fails to correctly predict the dryout heat flux due to the omission of thermal effects.
\begin{figure}[t!]
\includegraphics[scale=0.6]{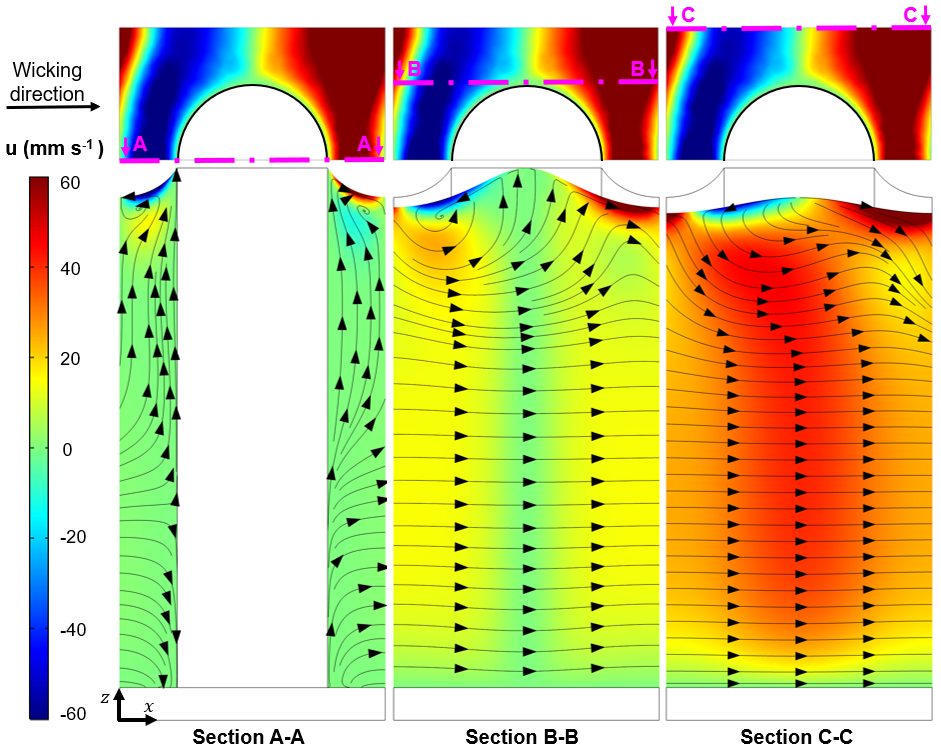}
\centering
\caption{Magnitude of the velocity component in the wicking ($x$-) direction with superimposed streamlines at different cross-sections of a unit-cell parallel to the wicking direction. Positions of cross-sections are (as specified in the top view): \textbf{a)} at the pillar center, \textbf{b)} at the pillar edge, and \textbf{c)} at the symmetry center between adjacent unit-cells. Results are given for Sample-8 experimented in Ref. \cite{wei_2018} at a heat load of $q''_ {in}$=50$\,  \rm W \, cm^{-2}$ for a unit-cell with the contact angle of $\theta$=30\degree.}
\label{fgr:Ma_Vel}
\end{figure}
A similar trend is also observed for the Marangoni model. Moderate variation of interfacial temperature keeps Ma number relatively small (see \cref{tab:case1table}), which leads to almost identical results for non-isothermal and Marangoni models in the first seven samples. Yet, for Sample-8, there is a slight deviation between the dryout heat flux predictions of non-isothermal and Marangoni models. Moreover, the prediction of the Marangoni model is closer to the experimental result, which gives a clue for the effect of thermocapillary convection. However, this slight difference prevents us from making a solid conclusion. Then the second comparison is made for the average evaporator superheats as shown in \cref{fgr:result_wei}b. Herein the isolated effect of thermocapillary convection is apparent. For Sample-8, the non-isothermal model fails to predict the average evaporator superheat, whereas the Marangoni model successfully predicts it. 
Since the average temperature of Sample-8 is substantially affected by thermocapillarity, Marangoni convection is expected to have a dominant effect on the liquid flow near the liquid-vapor interface. To reveal this, liquid flow patterns at different cross-sections of a unit-cell parallel to the wicking direction are presented in \cref{fgr:Ma_Vel} for Sample-8.
As shown in all cross-sections, thermocapillarity induces surface flows leading to circulations to a varying extent. While these surface flows are in the same direction with the wicking (capillary) flow at the downstream-half of a unit-cell, they oppose the wicking flow at the upstream-half of a unit-cell, which results in a stagnation region on the interface.  The origin of these thermocapillary surface flows is the interfacial temperature variation. Resultant temperature distributions on the interface (top view) together with different cross-sections parallel to the wicking direction are provided in \cref{fgr:Ma_temp}. As shown in the top view, a thermal boundary layer with a significant temperature gradient develops on the liquid-vapor interface near the pillar contact line. This temperature gradient triggers a radially outward surface flow from the warmer pillar contact line to the cooler zones around the pillar, leading to the flow patterns shown in \cref{fgr:Ma_Vel}.

The formation of periodic reverse surface flows, and the interfacial stagnation regions thereof, are expected to decrease the mass transport in the wicking direction. Accordingly, the dryout heat flux is expected to decrease due to the associated capillary flow deficit. The extent of the deficit, on the other hand, scales down with the pillar height since the permeability of a wick with substantially long pillars is slightly affected by the dynamics of the surface flows. The capillary flow through the pillars of \mbox{Sample-8}, for instance, exhibits an undisturbed parallel flow pattern in the wicking direction except for a restricted near interface zone affected by the thermocapillarity, which penetrates through the liquid film no more than a quarter of the pillar height (see \mbox{\cref{fgr:Ma_Vel}b-c}). Therefore, dryout heat flux is slightly affected by the presence of Marangoni flow in \mbox{Sample-8}. On the other hand, thermocapillary flow significantly enhances convective energy transport. As it can be seen in \cref{fgr:Ma_temp}, energy transport paths follow the velocity streamlines by manifesting the convection as the primary energy transport mechanism. The warmer liquid is moved away from the pillar contact line with the surface flow so that cooler fresh liquid from the outer region replenishes it. The resultant circulation creates an effective mixing mechanism that enhances the evaporation and the associated interfacial heat transfer coefficient. Accordingly, average evaporator temperature significantly drops \mbox{(\textit{ca.}$\,3 \,$\degree C)} in the presence of Marangoni convection.

\begin{figure}[b!]
\includegraphics[scale=0.55]{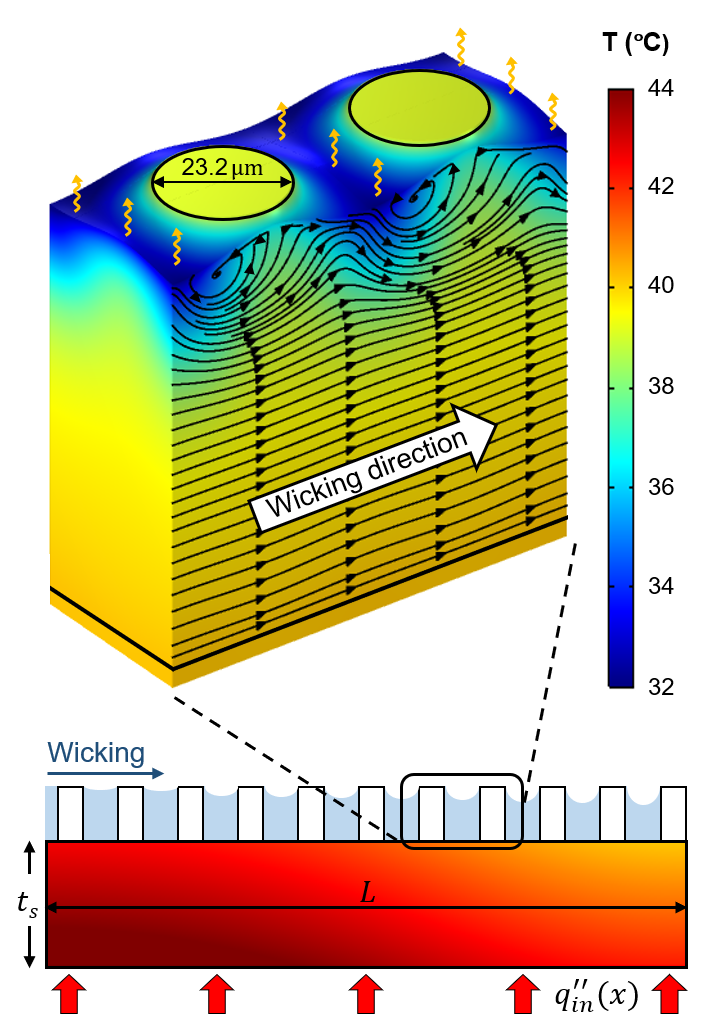}
\centering
\caption{The temperature distributions on the base wall of the substrate and two adjacent unit cells in the wicking direction. Velocity streamlines are superimposed on the cross-section parallel to the wicking direction.  Results are given for Sample-8 experimented in Ref. \cite{wei_2018} at a heat load of $q''_ {in}$=50$\,  \rm W \, cm^{-2}$ ($t_s$=570.3$\, \rm \mu m$, $L$=10$\, \rm mm$). Pillars shown on top of the base wall are drawn not to scale.}
\label{fgr:abs_imsge}
\end{figure}

A critical outcome of the model can be inferred from examining the 3-D temperature distribution on the evaporator device. Figure~\ref{fgr:abs_imsge} shows the temperature distribution on the base wall of the substrate together with the 3-D distribution of temperature in the vicinity of a pair of pillars. The nonuniform evaporation from the liquid-vapor interface results in a nonuniform temperature distribution on the base wall of the substrate. When a micropillar evaporator is used in an electronic cooling application such as chip cooling, the nonuniform temperature distribution on the evaporator may result in the development of on-chip thermal gradients, and undesired thermal stresses thereof.

\section{Conclusion}
\label{sec:conclusion}
Evaporation from a micropillar wick evaporator is modeled by coupling the capillary liquid flow with energy transfer in both liquid and solid domains. Thermocapillary flow is accounted for in the modeling for the first time in the literature. Predictions of the model are compared with a wide range of previous experimental results for water evaporation from micropillar evaporators, and excellent agreements are obtained. For the cases where the Ma number is relatively small, the effect of thermocapillarity on both dryout heat flux and average evaporator temperature is slight. However, when the Marangoni number is sufficiently high, thermocapillary convection sharply decreases the evaporator temperature by creating circulation beneath the liquid-vapor interface, which results in the formation of periodic reverse surface flows on the interface. This temperature reduction cannot be identified when thermocapillarity is not included in the modeling. Therefore, the current study successfully reveals the role of Marangoni flow in the evaporation of water from micropillar wick evaporators. We believe that our modeling approach can help researchers explore the relevant interfacial phenomena in the evaporation from the arrays of micro-scale surface structures and guide thermal scientists for the optimization of micro-post evaporators. 

\section*{Supplementary Material}
See Supplementary Material for mesh independence study, validation of the meniscus shape, bi-directional flow model, computational scheme, and assessment of the curvature effect.
\section*{Acknowledgements}
None.
\section*{Declaration of Interests}
\noindent The authors report no conflict of interest.
\addcontentsline{toc}{section}{References}
\bibliographystyle{unsrt}
\bibliography{references}
\end{document}